# A Dynamic Theory of Information and Entropy

**Michael C. Parker**

Fujitsu Laboratories of Europe, Columba House, Adastral Park, Ipswich, IP5 3RE, UK.

M.Parker@ftel.co.uk

**Stuart D. Walker**

University of Essex, Department of Electronic Systems Engineering, Wivenhoe Park, Colchester, Essex, CO4 3SQ, UK.

stuwal@essex.ac.uk

**Abstract**

**We propose a new thermodynamic, relativistic relationship between information and entropy, which is closely analogous to the classic Maxwell electro-magnetic equations. Determination of whether information resides in points of non-analyticity or is more distributed in nature therefore relates directly to the well-known wave-particle duality of light. At cosmological scales our vector differential equations predict conservation of information in black holes, whereas regular and Z-DNA correspond to helical solutions at microscopic levels. We further propose that regular and Z-DNA are equivalent to the alternative words chosen from an alphabet to maintain the equilibrium of an information transmission system.**

The close relationship between information and entropy is well recognised [1,2], e.g. Brillouin considered a negative change in entropy to be equivalent to a change in information (negentropy [3]), and Landauer considered the erasure of information $\Delta I$ to be associated with an increase in entropy $\Delta S$, via the relationship $-\Delta I = \Delta S$ [1,4]. In previous work, we qualitatively indicated the dynamic relationship between information and entropy, stating that the movement of information is accompanied by the increase of entropy in time [5,6], and suggesting that information propagates at the speed of light in vacuum $c$ [5-9]. Opinions on the physical nature of information have tended to be contradictory. One view is that information is inherent to points of non-analyticity (discontinuities) [5,7,9] that do not allow 'prediction', whereas other researchers consider the information to be more distributed (delocalised) in nature [10-12]. Such considerations are akin to the paradoxes arising from a wave-particle duality, with the question of which of the two complements best characterises information. In the following analysis, by initially adopting the localised, point of non-analyticity definition of information, we ultimately find that information and entropy do indeed also exhibit wavelike properties, and can be described by a pair of coupled wave equations, analogous to the Maxwell equations for electromagnetic (EM) radiation, and a distant echo of the simple difference relationship above. We start our analysis by consideration of a meromorphic [13] function $\psi(z)$ in a reduced 1+1 (i.e. one space dimension and one time dimension) complex space-time $z = x + ict$, with a simple, isolated pole at $z_0 = x_0 + ict_0$, as indicated in figure 1:

$$\psi(z) = \sqrt{\frac{ct_0}{\pi}} \frac{1}{z - z_0} \qquad (1)$$

The point of non-analyticity $z_0$ travels at the speed of light in vacuum $c$ [5-9]. We note that $\psi(z)$ acts as an inverse-square law function, e.g. as proportional to the field around an isolated charge, or the gravitational field around a point mass. It is also square-integrable, such that $\int_{-\infty}^{\infty} \rho|_{t=0} dx = 1$, and $ic\int_{-\infty}^{\infty} \rho|_{x=0} dt = 1$, where $\rho = |\psi(z)|^2$, and obeys the Paley-Wiener criterion for causality [14]. We now calculate the differential information [5,15] (entropy) of the function $\psi(z)$. By convention, in this paper, we define the logarithmic integration along the spatial $x$-axis to yield the entropy $S$ of the function; whereas the same integration along the imaginary $t$-axis yields the information $I$ of the function. This can be qualitatively understood to arise from the entropy of a system being related to its spatial permutations (e.g. the spatial fluxions of a volume of gas), whereas the information of a system is related to its temporal distribution (e.g. the arrival of a data train of light pulses down an optical fibre.) Considering the differential entropy first, substituting equation (1) with $t = 0$ and calculating the sum of the residues of the appropriate closed contour integral in the $z$-plane, $S$ is given by:

$$S = k \int_{-\infty}^{\infty} \rho \log_2 \rho \, dx = k \log_2 ct_0 \qquad (2)$$

where $k$ is the Boltzmann constant. We have defined $S$ using a positive sign in front of the integral, as opposed to the conventional negative sign [15], so that both the entropy and information of a space-time function are calculated in a unified fashion. In addition, this has the effect of aligning equation (2) with the 2$^{nd}$ Law of Thermodynamics, since $S$ increases monotonically with increasing time $t_0$. Performing the appropriate closed contour integral in the $z$-plane with $x = 0$ we find that the differential information is given by:

$$I = k \int_{-\infty}^{\infty} \rho \log_2 \rho \, cdt = k \log_2 x_0 . \qquad (3)$$

Note that in (2) and (3) some constant background values of integration have been ignored, since they disappear in the following differential analyses. We see that the entropy and information are given by surprisingly simple expressions. The information is the logarithm of the space-time distance $x_0$ of the point of non-analyticity from the temporal $t$-axis; whilst the entropy content is simply the logarithm of the distance (i.e. time $ct_0$) of the pole from the spatial $x$-axis. The overall info-entropy $\Sigma$ of the function $\psi(z)$ is given by the summation of the two orthogonal logarithmic integrals, $\Sigma = I + iS$, where the entropy $S$ is in quadrature to the information $I$, since $I$ depends on the real axis quantity $x_0$, whereas $S$ depends on the imaginary axis quantity $ct_0$. Since the choice of reference axes is arbitrary, we could have chosen an alternative set of co-ordinate axes to calculate the information and entropy. For example, consider a co-ordinate set of axes $x' - ct'$, rotated an angle $\theta$ about the origin with respect to the original $x - ct$ axes. In this case, the unitary transformation describing the position of the pole $z'_0$ in the new co-ordinate framework is:

$$\begin{pmatrix} ct'_0 \\ x'_0 \end{pmatrix} = \begin{pmatrix} \cos\theta & -\sin\theta \\ \sin\theta & \cos\theta \end{pmatrix} \begin{pmatrix} ct_0 \\ x_0 \end{pmatrix}. \qquad (4)$$

Using (2) and (3), the resulting values of the information and entropy for the new frame of reference are given by:

$$I' = k \log_2 x'_0 \qquad (5a) \qquad\qquad S' = k \log_2 ct'_0 \qquad (5b)$$

with overall info-entropy $\Sigma'$ of the function again given by the summation of the two quadrature logarithmic integrals, $\Sigma' = I' + iS'$. We next perform a dynamic calculus on the equations (5), with respect to the original $x - ct$ axes, using (4) to calculate:

$$\frac{\partial I'}{\partial x} = \frac{k}{x'_0}\frac{\partial x'_0}{\partial x} = \frac{k}{x'_0}\frac{\partial x'_0}{\partial x_0} = \frac{k}{x'_0}\cos\theta \qquad (6a) \qquad \frac{1}{c}\frac{\partial I'}{\partial t} = \frac{k}{x'_0}\frac{1}{c}\frac{\partial x'_0}{\partial t} = \frac{k}{x'_0}\frac{1}{c}\frac{\partial x'_0}{\partial t_0} = \frac{k}{x'_0}\sin\theta \qquad (6b)$$

$$\frac{1}{c}\frac{\partial S'}{\partial t} = \frac{k}{t'_0}\frac{1}{c}\frac{\partial t'_0}{\partial t} = \frac{k}{ct'_0}\frac{\partial t'_0}{\partial t_0} = \frac{k}{ct'_0}\cos\theta \qquad (7a) \qquad \frac{\partial S'}{\partial x} = \frac{k}{ct'_0}\frac{\partial t'_0}{\partial x} = \frac{k}{ct'_0}\frac{\partial t'_0}{\partial x_0} = \frac{-k}{ct'_0}\sin\theta \qquad (7b)$$

where we have ignored the common factor $\log 2$, and we have assumed the equality of the calculus operators $\partial/\partial x = \partial/\partial x_0$ and $\partial/\partial t = \partial/\partial t_0$, since the trajectory of the pole in the $z$-plane is a straight line with $dx = dx_0$, and $dt = dt_0$. Since the point of non-analyticity is moving at the speed of light in vacuum $c$ [5-9], such that $x_0 = ct_0$, we must also have $x'_0 = ct'_0$, as $c$ is the same constant in any frame of reference [16]. We can now see that equations (6) and (7) can be equated to yield a pair of Cauchy-Riemann equations [17]:

$$\frac{\partial I}{\partial x} = \frac{1}{c}\frac{\partial S}{\partial t} \qquad (8a) \qquad\qquad \frac{\partial S}{\partial x} = -\frac{1}{c}\frac{\partial I}{\partial t} \qquad (8b)$$

where we have dropped the primes, since equations (8) are true for all frames of reference. Hence, the info-entropy function $\Sigma$ is analytic (i.e. holographic in nature [13]), and the Cauchy-Riemann equations (8) indicate that information and entropy propagate as waves travelling at the speed $c$. In the Appendix we extend the analysis from 1+1 complex space-time to the full 3+1 (three-space and one-time dimensions) case, so that it can be straightforwardly shown that equations (8) generalise to:

$$\nabla \times \underline{I} = \frac{1}{c}\frac{d\underline{S}}{dt} \qquad (9a) \qquad\qquad \nabla \times \underline{S} = \frac{-1}{c}\frac{d\underline{I}}{dt} \qquad (9b)$$

where $\underline{I}$ and $\underline{S}$ are in 3D vector form. Further basic calculus manipulations reveal that the scalar gradient (divergence) of the information and entropy fields are both equal to zero:

$$\nabla \cdot \underline{I} = 0 \qquad (10a) \qquad\qquad \nabla \cdot \underline{S} = 0 \qquad (10b)$$

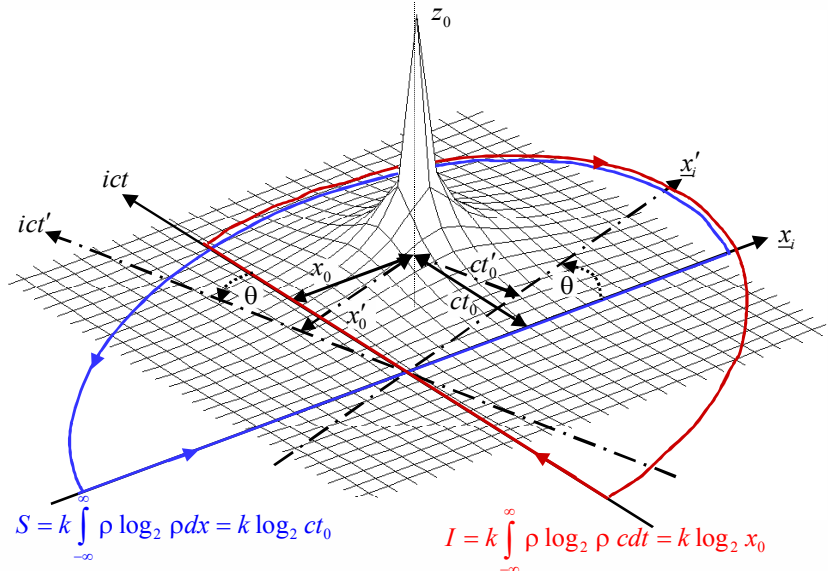

Figure 1: Information $I$ (integration over imaginary $ct$-axis) and entropy $S$ (integration over real $x$-axis) due to a point of non-analyticity (pole) at the space-time position $z_0$.

Together, equations (9) and (10) form a set of equations analogous to the Maxwell equations [16] (with no sources or isolated charges, and in an isotropic medium). Equations (9) can be combined to yield wave equations for the information and entropy fields:

$$\nabla^2 \underline{I} - \frac{1}{c^2}\frac{d^2\underline{I}}{dt^2} = 0 \qquad (11a) \qquad\qquad \nabla^2 \underline{S} - \frac{1}{c^2}\frac{d^2\underline{S}}{dt^2} = 0 \qquad (11b)$$

As defined previously, the information and entropy fields are mutually orthogonal, since $\underline{I} \cdot \underline{S} = 0$. Analogous to the Poynting vector describing energy flow of an EM wave, the direction of propagation of the wave is given by $\underline{I} \times \underline{S}$, e.g. see figure 2. The dynamic relationship between $I$ and $S$ means that information flow must be accompanied by entropy flux, such that movement of data is dissipative in order to satisfy the 2[nd] Law of Thermodynamics. Analytic functions such as $I$ and $S$ only have points of inflection, so that, again, the 2[nd] Law with respect to the $S$-field is a natural consequence of its monotonicity. However, in analogy to an EM-wave, equations (9) and (10) in combination imply that the sum of information and entropy $|\underline{I}|^2 + |\underline{S}|^2$ is a conserved quantity. Hence, the 2[nd] Law of Thermodynamics should be viewed as a conservation law for info-entropy,

similar to the 1st Law for energy. This has implications for the treatment of information falling into a black hole. The reason that the differential equations (6) and (7) obey the laws of relativity is that they are simple spatial reciprocal quantities, with their ratios equivalent to velocities, such that the relativistic laws are applicable. This reciprocal space aspect implies that calculation of Fourier-space info-entropy quantities result in quantities proportional to real space, which are therefore also relativistically-invariant and holographic.

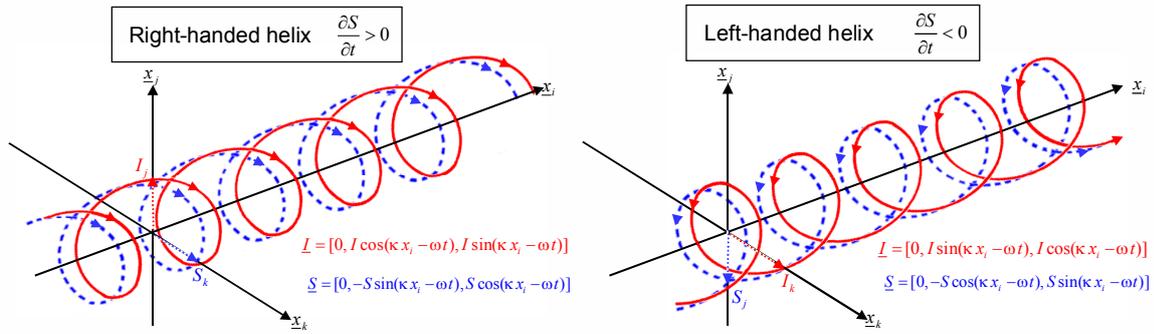

Figure 2: a) Right-handed polarisation helical info-entropy wave, propagating in positive $\underline{x}_i$-direction given by $\underline{I} \times \underline{S}$, b) Left-handed polarisation helical info-entropy wave travelling in same direction.

Equation (9a) shows that a right-handed helical spatial information distribution is associated with an increase of entropy with time (Fig 2a); in contrast to a left-handed chirality where entropy decreases with time (Fig 2b). A link is revealed between the high information density of right-handed DNA [18] and the overall increase of entropy with time in our universe [19]. A molecule such as DNA is an efficient system for information storage. However, our dynamic theory suggests that information would radiate away, unless a localisation mechanism were present, e.g. the existence of a standing wave. Such waves require forward and backward wave propagation of the same polarisation. We see that the complementary anti-parallel (C2 spacegroup) symmetry of DNA's double helix means that information waves of the same polarisation (chirality) will travel in both directions along the helix axis to form a standing wave, thus localising the information. Small left-handed molecules (e.g. amino acids, enzymes and proteins, as well as short sections of Z-DNA [20]) are known to exist alongside DNA within the cell nucleus. Their chirality is associated with a decrease of entropy with time. In conjunction with right-handed molecules they can be understood to regulate the thermodynamic function of the DNA molecule. As is well known, highly dynamic systems require negative feedback to direct and control their output, without which they become unstable. We can draw parallels to the high entropic potential of DNA, with left-handed molecules (as well as right-handed molecules) acting as damping agents to control its thermodynamic action.

In conclusion, our theory grounds information and entropy onto a physical, relativistic basis, and provides thermodynamic insights into the double-helix geometry of DNA. It offers the prediction that non-DNA-based life forms will still tend to have helical structures, and suggests that black holes conserve info-entropy.

**Appendix**

We generalise the $x$-dimension to the $i^{\text{th}}$ space-dimension ($i = 1, 2, 3$), such that for a pole travelling in the $\underline{x}_i$-direction (i.e. equivalent to a plane wave travelling in the $\underline{x}_i$-direction) the info- and entropy-fields vibrate in the mutually-orthogonal $\underline{x}_j$- and $\underline{x}_k$-directions respectively, where again $j, k = 1, 2, 3$ and $i \neq j \neq k$. The vector descriptions of the $\underline{I}$ and $\underline{S}$ fields are:

$$\underline{I} = I_1 \underline{x}_1 + I_2 \underline{x}_2 + I_3 \underline{x}_3 \quad \text{(A1a)} \qquad \underline{S} = S_1 \underline{x}_1 + S_2 \underline{x}_2 + S_3 \underline{x}_3 \quad \text{(A1b)}$$

In analogy to an EM plane-wave, two plane-wave polarisations are therefore possible:

(a) $\quad I_i = 0$, $I_j = k \log_2 x'_i$, $I_k = 0 \qquad S_i = 0$, $S_j = 0$, $S_k = k \log_2 ct'$ (A2a)

(b) $\quad I_i = 0$, $I_j = 0$, $I_k = -k \log_2 x'_i \qquad S_i = 0$, $S_j = k \log_2 ct'$, $S_k = 0$ (A2b)

where we have generalised the position of the point of non-analyticity (pole) to a position $z' = x'_i + ict'$, i.e. dropped the subscript '0'. We see that the information and entropy fields are mutually orthogonal, since $\underline{I} \cdot \underline{S} = 0$. Also, the direction of propagation of the wave is given by $\underline{I} \times \underline{S}$ with the sign of $I_{j,k}$ chosen so that flow is in the positive $\underline{x}_i$-direction. We make use of the $4 \times 4$ transformation matrix $\underline{g}$, relating one relativistic frame of reference to another [16]:

$$\begin{pmatrix} ict' \\ x'_1 \\ x'_2 \\ x'_3 \end{pmatrix} = \begin{pmatrix} \gamma & -\gamma\beta_1 & -\gamma\beta_2 & -\gamma\beta_3 \\ -\gamma\beta_1 & 1+A\beta_1^2 & A\beta_1\beta_2 & A\beta_1\beta_3 \\ -\gamma\beta_2 & A\beta_2\beta_1 & 1+A\beta_2^2 & A\beta_2\beta_3 \\ -\gamma\beta_3 & A\beta_3\beta_1 & A\beta_3\beta_2 & 1+A\beta_3^2 \end{pmatrix} \begin{pmatrix} ict \\ x_1 \\ x_2 \\ x_3 \end{pmatrix} \quad \text{(A3)}$$

where $\beta_1$, $\beta_2$, and $\beta_3$ are the normalised velocities (with respect to $c$) in the co-ordinate directions, with $\beta^2 = \beta_1^2 + \beta_2^2 + \beta_3^2 = 1$ in this case, since the overall wave velocity is equal to $c$. In addition the parameters $\gamma$ and $A$ are related by $\gamma = 1/\sqrt{1-\beta^2}$, and also $\gamma = 1 + A\beta^2$. We note that equation (A3) is made equivalent to (4), by substituting $\gamma = \cosh i\theta$, and $\beta_1 = \tanh i\theta$, with $\beta_2 = \beta_3 = 0$. Performing a dynamic calculus on equations (A2a), we derive the following expressions:

$$\frac{\partial I_j}{\partial x_i} = \frac{k}{x_i'}\frac{\partial x_i'}{\partial x_i} = \frac{k}{x_i'}g_{ii} \quad \text{(A4a)} \qquad \frac{\partial S_k}{\partial ict} = \frac{k}{ict'}\frac{\partial ict'}{\partial ict} = \frac{k}{ict'}g_{00} \quad \text{(A4b)}$$

$$\frac{\partial I_j}{\partial ict} = \frac{k}{x_i'}\frac{\partial x_i'}{\partial ict} = \frac{k}{x_i'}g_{i0} \quad \text{(A4c)} \qquad \frac{\partial S_k}{\partial x_i} = \frac{k}{ict'}\frac{\partial ict'}{\partial x_i} = \frac{k}{ict'}g_{0i} \quad \text{(A4d)}$$

Consideration of the second polarisation (b) allows us to derive the additional following expressions:

$$\frac{\partial I_k}{\partial x_i} = \frac{-k}{x_i'}\frac{\partial x_i'}{\partial x_i} = \frac{-k}{x_i'}g_{ii} \quad \text{(A5a)} \qquad \frac{\partial S_j}{\partial ict} = \frac{k}{ict'}\frac{\partial ict'}{\partial ict} = \frac{k}{ict'}g_{00} \quad \text{(A5b)}$$

$$\frac{\partial I_k}{\partial ict} = \frac{-k}{x_i'}\frac{\partial x_i'}{\partial ict} = \frac{-k}{x_i'}g_{i0} \quad \text{(A5c)} \qquad \frac{\partial S_j}{\partial x_i} = \frac{k}{ict'}\frac{\partial ict'}{\partial x_i} = \frac{k}{ict'}g_{0i} \quad \text{(A5d)}$$

Since the direction of propagation is in the $x_i$-direction, we must have that $\beta_j = \beta_k = 0$, so that $\beta_i = \beta$, and therefore $\gamma = 1 + A\beta_i^2$, such that $g_{ii} = g_{00}$, as well as $g_{i0} = g_{0i}$. Given $x_i' = ct'$ as usual, in the $x_i$-direction the following pair of equations holds:

$$-\frac{\partial I_k}{\partial x_i} = \frac{1}{c}\frac{\partial S_j}{\partial t} \quad \text{(A6a)} \qquad -\frac{\partial S_k}{\partial x_i} = -\frac{1}{c}\frac{\partial I_j}{\partial t} \quad \text{(A6b)}$$

Likewise, in the $x_k$-direction we find the following pair of equations holds:

$$\frac{\partial I_j}{\partial x_i} = \frac{1}{c}\frac{\partial S_k}{\partial t} \quad \text{(A7a)} \qquad \frac{\partial S_j}{\partial x_i} = -\frac{1}{c}\frac{\partial I_k}{\partial t} \quad \text{(A7b)}$$

We see that both pairs of equations (A6) and (A7) respectively obey the Cauchy-Riemann symmetries evident in equations (8). Considering the three cyclic permutations of $i, j$ and $k$, we can write:

$$\sum_{i=1}^{3}\left(\frac{\partial I_j}{\partial x_k} - \frac{\partial I_k}{\partial x_j}\right)x_i = \frac{1}{c}\sum_{i=1}^{3}\frac{\partial S_i}{\partial t}x_i \quad \text{(A8a)} \qquad \sum_{i=1}^{3}\left(\frac{\partial S_j}{\partial x_k} - \frac{\partial S_k}{\partial x_j}\right)x_i = \frac{-1}{c}\sum_{i=1}^{3}\frac{\partial I_i}{\partial t}x_i \quad \text{(A8b)}$$

Equations (A8) can be alternatively written in the compact vector notation of equations (9).

**References**

[1] C.H. Bennett, Scientific American, **257**, 88 (1987)

[2] R. Landauer, Physica Scripta, **35**, 88 (1987)

[3] L. Brillouin, *Science & Information Theory*, (Academic, New York, 1956)

[4] R. Landauer, Physics Today, **44**, 23 (1991)

[5] M.C. Parker and S.D. Walker, Optics Communications, **229**, 23 (2004)

[6] M.C. Parker and S.D. Walker, http://arxiv.org/abs/physics/0401077 (2004)

[7] J.C. Garrison, M.W. Mitchell, R.Y. Chiao, and E.L. Bolda, Physics Letters A, **245**, 19 (1998)

[8] L. Brillouin, *Wave propagation and group velocity*, (Academic, New York, 1960)

[9] M.D. Stenner, D. J. Gauthier, and M.A. Neifeld, Nature, **425**, 695 (2003)

[10] K. Wynne, Optics Communications, **209**, 85 (2002)

[11] J.J. Carey, J. Zawadzka, D. A. Jaroszynski, and K. Wynne, Physical Review Letters, **84**, 1431 (2000)

[12] W. Heitmann, G. Nimtz, Physics Letters A, **196**, 154 (1994)

[13] G.B. Arfken, H.J. Weber, *Mathematical Methods for Physicists*, (Academic, New York, 1995), Chap. 6

[14] H. Primas, *Time, Temporality, Now: The representation of facts in physical theories*, (Springer, Berlin, 1997) p.241-263

[15] N. Gershenfeld, *The physics of information technology*, (Cambridge University Press, Cambridge, 2000), Chap. 4

[16] J.D. Jackson, *Classical Electrodynamics*, (John Wiley & Sons, New York, 1999), Chap. 11

[17] K.-E. Peiponen, E.M. Vartianen, and T. Asakura, *Dispersion, complex analysis, and optical spectroscopy: classical theory*, (Springer, Berlin, 1999)

[18] J.D. Watson and F.H.C. Crick, Nature, **171**, 737 (1953)

[19] W.H. Zurek, *Complexity, entropy and the physics of information*, (Addison-Wesley, Redwood City, 1989), Vol. 8

[20] J.S. Siegel, Nature, **409**, 777 (2001)